\documentclass[12pt]{iopart}
\usepackage{iopams}
\usepackage{graphicx}
\begin{document}
\title[]{Introducing the Hamiltonian as a ``thermodynamic'' potential}

\author{V F Correa}

\address{Centro At\'omico Bariloche (CNEA) and Instituto Balseiro (U. N. Cuyo), 8400 Bariloche, R\'io Negro, Argentina}

\ead{victor.correa@cab.cnea.gov.ar}

\vspace{10pt}
\begin{indented}
\item[] December 2018
\end{indented}

\begin{abstract}
A conceptually simple physical interpretation of a conserved Hamiltonian $\mathcal{H}$ for a mechanical system with a time-dependent constraint is given. For the case of a bead on a vertical hoop forced to rotate with constant angular velocity $\omega$, $\mathcal{H}$ is nothing but the total energy of the system plus the external actuator keeping $\omega$ fixed. In an analogy with thermodynamics, the Hamiltonian is introduced as a thermodynamic potential obtained from a Legendre transformation of the energy, in a very instructive way. The ideas can be made extensive to different problems with time-dependent constraints.
\end{abstract}

\vspace{2pc}

\submitto{\EJP}
\maketitle

\normalsize

\section{Introduction}

When teaching the Lagrange formalism in undergraduate or graduate classical mechanics courses it is usual to deal with problems in which the mechanical energy $E_M$ is not conserved, but the Hamiltonian $\mathcal{H}$ is. While this last property is a consequence that the Lagrangian $\mathcal{L}$ does not depend explicitly on time, the time dependent energy is due to the presence of non-conservative forces. In these situations, it is usual to hear the same question from students: so, if $\mathcal{H}$ is constant, what is it? what is its physical meaning?
There is no general answer to these questions. Nonetheless, we give below an example of a standard non-conservative problem where $\mathcal{H}$ has a straightforward physical interpretation. We show as well that $\mathcal{H}$ can be obtained from a Legendre transformation of $E_M$ (besides its usual definition as a Legendre transformation of $\mathcal{L}$) resulting in a simple and interesting parallelism between $\mathcal{H}$ and a thermodynamic potential. This perspective can be useful to help students understand the physics behind the Hamiltonian. 

\section{A bead on a rotating hoop}

\begin{figure}[t]
\centering
\includegraphics[width=0.4\columnwidth]{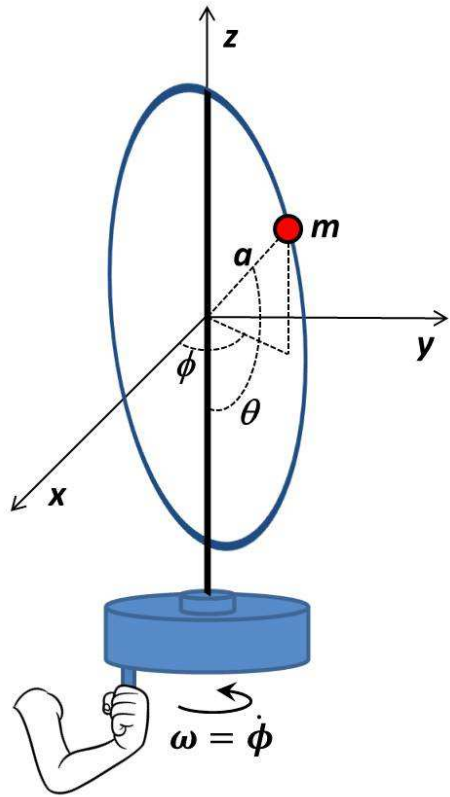}
\caption[]{(color online) A bead of mass $m$ on a rotating hoop of radius $a$ and constant angular velocity $\omega$, under gravity.}
\label{fig1}
\end{figure}

Let us consider the very well known mechanical setup depicted in Fig. \ref{fig1}: a bead of mass $m$ on a rotating massless hoop of radius $a$ that is externally constrained to move with constant angular velocity $\omega$ around the vertical $\hat{z}$ axis. Due to this constraint, this is a problem with just one degree of freedom. The best choice of a generalized coordinate in order to write down $\mathcal{L}$ is the angular variable $\theta$ as shown in Fig. \ref{fig1}. The problem is treated in many textbooks \cite{Taylor2005,Morin2008,Ponce2010} as an example of time-dependent constraints (i.e., $\phi = \omega t$). 

Due to the gravitational force acting on $m$, there is a critical frequency $\omega_c = \sqrt{g/a}$ below which the only stable equilibrium occurs at $\theta = 0$. For $\omega > \omega_c$, the stable equilibrium $\theta_o$ is given by $\cos \theta_o = g / a \omega^2$, with $\theta_o$ continuously moving from 0 to $\pi / 2$ as $\omega$ increases. Much has been said and written about the many different aspects of this problem and slight variations of it \cite{Raviola2017}. 
However, we are not interested here in the detailed solution nor in the dynamics of the system. Instead we are going to focus on more general matters, mainly associated with the conserved quantities.

The Lagrangian $\mathcal{L} = T - U$ of this 1-degree of freedom system is

\begin{equation}
\mathcal{L} (\theta,\dot{\theta}) = \frac{1}{2} m a^2 (\dot{\theta}^2 + \omega^2 \sin^2 \theta) + m g a \cos \theta 
\label{Lw}
\end{equation}

\noindent The mechanical energy $E_M = T + U$ is not a conserved quantity (i.e., $dE_M / dt \neq 0$) since an external actuator is doing work on the system to keep $\omega$ fixed. The system is non-conservative not in the sense that there are dissipative forces, but in the sense that there are forces doing work that have not been derived from a conservative potential energy. In this case, the external force acting to keep $\omega$ fixed. 

The Hamiltonian, on the other hand, calculated from its definition as the Legendre transformation of $\mathcal{L}$

\begin{equation}
\mathcal{H} = \frac{\partial \mathcal{L}}{\partial \dot{\theta}} \dot{\theta} - \mathcal{L} = 
\frac{1}{2} m a^2 (\dot{\theta}^2 - \omega^2 \sin^2 \theta) - m g a \cos \theta 
\label{H}
\end{equation}

\noindent is conserved.\footnote{To be precise, $\mathcal{H}$ should be written explicitly in terms of its natural independent variables, $\theta$ and its conjugated momentum $p_{\theta}= \frac{\partial \mathcal{L}}{\partial \dot{\theta}}$. But we are not interested in these ``minor'' details here.}
Since it is conserved, does it have any physical interpretation? Let us see.

The external constraint torque that keeps $\omega$ constant is collinear with the vertical rotation axis of the hoop ($\hat{z}$ axis). Any other torque component is canceled by the constraint that the hoop only rotates around the vertical axis. We can calculate this torque from its dynamical effects as $N_z = \frac{d L_z}{dt}$. The angular momentum is given by $L_z = I_z \omega$ where $I_z = m a^2 \sin^2 \theta$ is the moment of inertia around the $\hat{z}$ axis.
The work done by the torque is

\begin{eqnarray}
W &=& \int^{\phi}_{0} N d\phi' =  \int^{\phi}_{0} \frac{dL_z}{dt} d\phi' = \int^{L_z}_{0} dL_z' \frac{d\phi'}{dt}  = 
\int^{L_z}_{0} dL_z' \omega  = \omega L_z \nonumber \\
\nonumber \\
  &=& m \omega^2 a^2 \sin^2 \theta 
\label{N}
\end{eqnarray}  

\noindent since $\omega$ is constant.\footnote{The lower limit of the integral in $L_z$ has been arbitrarily set to 0. Any other value would just only change the arbitrary energy zero, which has no physical significance.} This work is done by the external actuator. Thus, the energy change in the actuator after doing this work is $E_a = - W$. The total energy of the bead-hoop system plus the external actuator is then 

\begin{eqnarray}
E_T &=& E_m + E_a   \\ 
    &=& \frac{1}{2} m a^2 (\dot{\theta}^2 + \omega^2 \sin^2 \theta) - m g a \cos \theta - m a^2 \omega^2 \sin^2 \theta  \nonumber \\
		&=& \frac{1}{2} m a^2 (\dot{\theta}^2 - \omega^2 \sin^2 \theta) - m g a \cos \theta = \mathcal{H}  \nonumber
\label{ET}
\end{eqnarray} 

So, we arrive at the very interesting result that the conserved Hamiltonian is nothing but the total energy of the original system plus the actuator, that is, of the system and its surroundings, which obviously must be a conserved quantity, unless dissipative forces were acting (which are not).

The mechanical energy $E_M$ is not conserved because in the Lagrange formalism the work done by the constraint forces is not included. When these forces are associated with time-independent constraints, they do not do work. So, not including them has no effect and $E_M$ is conserved unless dissipative forces are acting. But when the constraint forces are linked to time-dependent constraints (as in our case), they indeed do work and because this work is not included in $E_M$, it cannot conserve \cite{Goldstein1950}.

\section{Hamiltonian and thermodynamic potentials}

We are going to make use of the fact that $\mathcal{H}$ is a conserved energy to go a step forward and make an interesting analogy between $\mathcal{H}$ and the thermodynamic potentials.
If the constraint $\dot{\phi} = \omega$ is relaxed, the system bead + hoop has now two degrees of freedom. Choosing $\theta$ and $\phi$ as generalized coordinates (see Fig. \ref{fig1}), the Lagrangian is written as 

\begin{equation}
\mathcal{L} (\theta,\dot{\phi},\dot{\theta}) = \frac{1}{2} m a^2 (\dot{\theta}^2 + \dot{\phi}^2 \sin^2 \theta) + m g a \cos \theta. 
\label{L}
\end{equation}  

\noindent The energy is conserved now since the actuator is not working. On the other hand, $\phi$ is a cyclic coordinate. Hence, its conjugate momentum is also conserved. This momentum is just the $\hat{z}$-component of the angular momentum, $L_z = m \dot{\phi} a^2 \sin^2 \theta$. The energy is given by 

\begin{equation}
E_M(\theta,\dot{\theta},L_z) = \frac{1}{2} m a^2 \dot{\theta}^2 + \frac{1}{2} \frac{L_z^2}{m a^2 \sin^2 \theta} - m g a \cos \theta 
\label{Ephi}
\end{equation} 

\noindent where the first term and the last two terms to the right correspond to the kinetic energy and the potential energy of the effective one dimensional reduced problem, respectively. Since $E_M$ is conserved, Eqn. \ref{Ephi} is the ``correct potential'' and, from its minimization (i.e., $\left(\frac{\partial E_M}{\partial \theta}\right)_{L_z} = 0$), one can obtain the equilibrium states of the bead. 

The equilibrium state of a thermodynamic system is ruled by the second law. If, for simplicity, we consider a simple system whose equation of state can be described by just the volume, temperature and pressure, the $2^{nd}$ law states that in equilibrium the internal energy $U$ of such a system is a minimum if the entropy and the volume are kept constant. This formulation of the $2^{nd}$ law is sometimes referred as the Energy Minimum Principle (EMP) \cite{Callen1985} and is very useful for our purposes. 

In the real world it is hard to find any process occurring at constant values of extensive parameters like the entropy or volume. Instead, most processes occur at a constant value of an intensive parameter, for instance, the pressure $P$. The concept of a reservoir is useful in these cases \cite{Callen1985}. For instance, a $P$-reservoir is a system so large that any volume change affecting it can be considered negligible in such a way that the pressure inside it is always constant. If it is put in contact with other system, it also fixes the pressure of this system. A $P$-reservoir only exchanges energy in the form of mechanical work $PdV$. 

So, let us put our simple thermodynamic system in contact with a $P$-reservoir. The equilibrium state can no longer be obtained from the minimization of $U$ since $V$ is not constant. In order to apply correctly the EMP we need to consider the whole ``isolated'' arrangement: system + $P$-reservoir, whose total volume (and entropy) is constant. Sometimes, this can be a difficult task. But the magnificence of the thermodynamics comes to our help and establishes that the equilibrium state is achieved from the potential obtained from a Legendre transformation of $U$ with respect to $V$ (conjugate variable associated with $P$). In other words, minimizing $U_T = U + U_R$ is absolutely equivalent to minimize

\begin{equation}
H(S,P) = U(S,V) - \left(\frac{\partial U}{\partial V}\right)_S V = U(S,V) + PV 
\label{Hent}
\end{equation}

\noindent where the potential $H(S,P)$ is nothing but the enthalpy of the system. Thus, when the system is in contact with a $P$-reservoir, $H$ is the correct potential to obtain the equilibrium state, not $U$. Obviously, it is straightforward to show that $H(S,P) = U_T$. 

Let us go back to our mechanical bead+hoop system and make an analogy. If you put it in contact with a $\omega$-reservoir (i.e., a very powerful external actuator that keeps $\omega$ constant regardless what the bead does), the equilibrium state can no longer be obtained from the minimization of $E_M$ since $L_z$ is no longer conserved. We need to consider, in turn, the whole ``isolated'' arrangement: system + $\omega$-reservoir, whose total $L_z$ is constant. We have already seen that $E_M$ + $E_a$ is just $\mathcal{H}$. Anyway, let us follow the protocol and calculate the Legendre transformation of $E_M$ with respect to $L_z$   

\begin{eqnarray}
E_M((\theta,\dot{\theta},L_z)) - \left(\frac{\partial E_M}{\partial L_z}\right) L_z &=& E_M((\theta,\dot{\theta},L_z)) - \omega L_z  \nonumber \\ 
		&=& \frac{1}{2} m a^2 (\dot{\theta}^2 - \omega^2 \sin^2 \theta) - m g a \cos \theta    \\
		&=& \mathcal{H}(\theta,\dot{\theta},\omega)  \nonumber
\label{Htrans}
\end{eqnarray} 

\noindent So, $\mathcal{H}$ is the correct ``potential'' to analize the equilibrium state when the system is in contact with a $\omega$-reservoir.

\section{Conclusions}

Making use of a simple mechanical problem, i.e. a massive bead on a massless rotating hoop at constant $\omega$, we analyze the physics behind the conserved Hamiltonian. We show that it corresponds to the total energy of the bead-hoop system plus the external actuator that keeps $\omega$ fixed. We also present an interesting and instructive analogy between the Hamiltonian and the thermodynamic potentials that can help students to understand its physical meaning.

\section{Acknowledgments}

The author thanks J. J. Guimpel and G. Abramson for a careful and critic reading of the manuscript. V. F. C is member of CONICET, Argentina.


\section*{References}

\begin{thebibliography}{10}

\bibitem{Taylor2005} Taylor J R 2005 {\it Classical Mechanics} (Sausalito, Calif: University Science Books) p~260

\bibitem{Morin2008} Morin D J 2008 {\it Introduction to Classical Mechanics with Problems and Solutions} (Cambridge, UK: Cambridge University Press) p~248, 264

\bibitem{Ponce2010} Ponce V H 2010 {\it Mec\'anica Cl\'asica} (Mendoza, Argentina: EDIUNC) p~75

\bibitem{Raviola2017} Raviola L A, V\'eliz M E, Salomone H D, Olivieri N A and Rodr\'iguez E E 2017 The bead on a rotating hoop revisited: an unexpected resonance \textit{Eur. J. Phys.} \textbf{38} 015005 and references therein 

\bibitem{Goldstein1950} Goldstein H 1950 {\it Classical Mechanics} (Reading, Mass: Addison-Wesley) p~54

\bibitem{Callen1985} Callen H B 1985 {\it Thermodynamics and an Introduction to Thermostatistics} (New York: John Wiley $\&$ Sons) ch~5, 6

  

\end{thebibliography}
\end{document}